\documentclass[%
 reprint,
superscriptaddress,
 amsmath,amssymb,
 aps,
]{revtex4-2}

\usepackage{graphicx}
\usepackage{dcolumn}
\usepackage{bm}
\usepackage{xcolor} 
\usepackage{hyperref}
\usepackage{upgreek}

\begin{document}

\preprint{APS/123-QED}

\title{Single-spin spectroscopy of spontaneous and phase-locked spin torque oscillator dynamics}
 
\author{Adrian Solyom}
\affiliation{%
Department of Physics, McGill University, 3600 Rue University, Montreal QC, H3A 2T8, Canada
}%
\author{Michael Caouette-Mansour}
\affiliation{%
Department of Physics, McGill University, 3600 Rue University, Montreal QC, H3A 2T8, Canada
}%
\author{Brandon Ruffolo}
\affiliation{%
Department of Physics, McGill University, 3600 Rue University, Montreal QC, H3A 2T8, Canada
}%
\author{Patrick Braganca}
\affiliation{%
Western Digital, San Jose CA, 95119 USA
}%
\author{Lilian Childress}%
\affiliation{%
Department of Physics, McGill University, 3600 Rue University, Montreal QC, H3A 2T8, Canada
}%
\author{Jack Sankey}
 \email{jack.sankey@mcgill.ca}
\affiliation{%
Department of Physics, McGill University, 3600 Rue University, Montreal QC, H3A 2T8, Canada
}%

\date{\today}

\begin{abstract}
We employ N-$V$ magnetometry to measure the stray field dynamics of a ferromagnetic permalloy nanowire driven by spin-orbit torques. Specifically, we observe the optically detected magnetic resonance (ODMR) signatures of both spontaneous DC-driven magnetic oscillations and phase-locking to a second harmonic drive, developing a simple macrospin model that captures the salient features. We also observe signatures of dynamics beyond the macrospin model, including an additional ODMR feature (associated with a second SW mode) and one mode sapping power from another. Our results provide additional insight into N-$V$-spin wave coupling mechanisms, and represent a new modality for sub-wavelength N-$V$ scanned probe microscopy of nanoscale magnetic oscillators.
\end{abstract}

\maketitle

\section{Introduction}

When compared to architectures employing field-effect transistors, logical devices using spin degrees of freedom to store and carry information show promise for energy-efficient processing \cite{Kruglyak2010, Chumak2014, Chumak2015, WangQ2020}. Moreover, the non-linear nature of such magnetic devices offers potential new avenues for massively parallel neuromorphic computing \cite{Lequeux2016, Romera2018, Shibata2020, Siddiqui2020, Grollier2020, Yang2021, Schuman2022}. The development of this new technology requires characterization methods appropriate for a range of devices, such as logic gates \cite{WangQ2020} and oscillator arrays \cite{Zahedinejad2021}. Established techniques such as magnetoresistive readout \cite{Kiselev2003, Kiselev2005, Sankey2006, Liu2011, Duan2014}, Brillouin light scattering \cite{Demokritov2001, Demokritov2007, Serga2010}, x-ray scattering \cite{Ament2011, Trager2020}, and time-resolved magneto-optic Kerr effect measurements \cite{Kruglyak2010, Au2011, Bauer2014} allow for excited magnetic spin-waves (SWs) to be detected and measured in frequency space, but each method naturally causes backaction on the device and / or is limited in spatial resolution. To address these two issues, solid state spins have been developed as atomic-scale, minimally invasive probes of the stray fields surrounding SW excitations. Among spin-based magnetometers, the nitrogen-vacancy (N-$V$) defect in diamond is currently the most studied, and can be placed within nanometers of the sample, enabling scanned probe readout of SWs with small wavelengths. To date, N-$V$s have been used to probe ferromagnetic phenomena, including vortex cores \cite{Rondin2013, Wolf2016, Scheidegger2022}, domain walls \cite{Tetienne2015}, oscillators \cite{Zhang2020}, magnetic tunnel junctions \cite{Yan2022}, SW dispersion \cite{Bertelli2020, Simon2022}, and SW scattering \cite{Gonzalez2022}.

In this study, we employ N-$V$ magnetometry to measure the stray field of a metallic ferromagnetic nanowire driven by DC and microwave spin-orbit torques (plus associated magnetic fields) generated by current in a platinum capping layer. Following Refs. \cite{Solyom2018, Lee-Wong2020, Simon2022}, we parametrically drive SW modes at twice their natural frequency, directly observing the response in the N-$V$'s optically detected magnetic resonance (ODMR) spectrum at the SW frequency (without directly driving the N-$V$ transistions). To extend previous work, we design the present device with dimensions small enough to suppress Suhl instabilities \cite{Suhl1957, Haghshenasfard2017}, allowing it to act as a spin-torque oscillator (STO) with large-amplitude SW oscillations driven by DC bias alone. Below the STO threshold current, we observe parametrically driven large-angle precession in the ODMR spectrum, and resolve two SW modes, one of of which produces a signature of auto-oscillation above the threshold current. In this regime, the ODMR spectrum also exhibits evidence of coupling between the modes, as parametrically driving the second saps power from the the first. Furthermore, we identify the ODMR spectral signature of phase locking between auto-oscillations and a second-harmonic drive by comparing our ODMR spectra with a simple model coupling a macrospin \cite{Solyom2018} to an N-$V$, and find semi-quantitative agreement with the behavior of the fundamental SW mode. Finally, we perform spin relaxometry with the N-$V$ to observe the magnetic noise from the SW modes as the bias is increased above the STO threshold. The results from this simple testbed system demonstrate the potential utility of N-$V$ magnetometry in STO characterization that should be especially useful when applied in a scanned-probe measurement \cite{Degen2008, Maletinsky2012, Simon2021, Zhou2021, Simon2022}.

\section{Device fabrication and transport characterization}

\begin{figure}[ht]
\includegraphics[width=8.5cm]{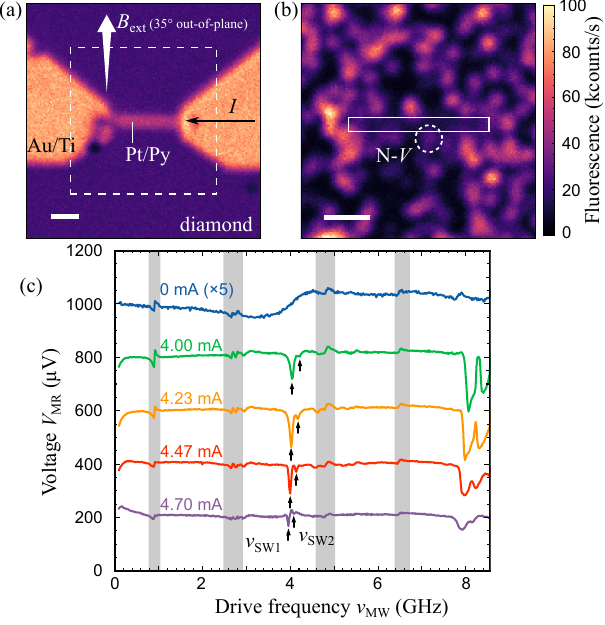}
\caption[Device geometry and transport measurements]{
\label{fig1:dev-transport} 
\textbf{Device geometry and transport measurements.} 
(a) Reflected confocal image of device with overlapping leads attached. The dashed box indicated the region imaged in (b). (b) Fluorescence confocal image of N-$V$ centers near the device, taken prior to the deposition of electrical leads in (a). The white rectangle indicates the location of the device (a shadow in fluorescence), while the dotted circle highlights the location of the N-$V$ center used for measurements in this study. Scale bars in (a) and (b) are 1 $\upmu$m each. (c) ST-FMR measurements performed with $B_\mathrm{ext} = 41$ mT, $I_\mathrm{MW} = 1$ mA, and with varied $I_\mathrm{bias}$. Arrows mark the narrow synchronized response of the first (second) STO modes at $\nu_{\mathrm{SW}1}$ ($\nu_{\mathrm{SW}2}$). The gray shaded regions indicate ``untrustworthy'' frequency ranges with low transmission to the device. Data are offset vertically for clarity, and the zero-bias trace is scaled by a factor of 5 to enhance visibility.}
\end{figure}

We study Py(Ni$_{80}$Fe$_{20}$, 5 nm)/Pt(5 nm) nanowires with nominal lateral dimensions $3$ $\upmu$m $\times~0.3$ $\upmu$m fabricated on an electronic-grade diamond substrate (see Appendix \ref{subsection:Device fab} for fabrication details). Figure \ref{fig1:dev-transport}(a) shows an optical micrograph of the device after the electrical leads have been deposited with a 300-nm overlap on either side of the nanowire. The dashed area in Fig. \ref{fig1:dev-transport}(b) is a confocal fluorescence image taken (by focusing green 532-nm light to a scanned point and collecting red fluorescence with the same objective) before depositing the leads. Individual N-$V$ centers appear as bright spots. The device, which blocks light from entering or leaving the diamond substrate, appears as a shadow highlighted with a white rectangle, and the N-$V$ used to probe the stray field is located within the dotted white circle. Throughout this study, a magnetic field is applied along the diamond's [111] direction, which is orthogonal to the current flow in the nanowire and parallel to the N-$V$ symmetry axis (canted approximately 35 degrees out of plane) \cite{Solyom2018}.

We electrically characterize the magnetic modes by performing spin-transfer ferromagnetic resonance (ST-FMR), using a pulse-modulated lock-in technique \cite{Solyom2018}. The measurement works by detecting the average voltage $V_{\mathrm{MR}}$ resulting from the mix-down between an applied microwave (MW) current and magnetoresistance oscillations associated with magnetic precession (averaged over the volume of the magnetic layer). Figure \ref{fig1:dev-transport}(c) shows such measurements performed on the device while sweeping the drive frequency $\nu_{\mathrm{MW}}$ and applying a magnetic field of $B_{\mathrm{ext}} = 41$ mT with current $I(t) = I_{\mathrm{MW}} \cos( 2 \pi \nu_{\mathrm{MW}} t) + I_{\mathrm{bias}}$ comprising microwave drive amplitude $I_{\mathrm{MW}} = 1$ mA and a dc bias $I_{\mathrm{bias}}$. At zero bias (top blue curve, scaled by factor of 5 for clarity), we observe a characteristic Fano line shape at 4 GHz, as expected for ferromagnetic resonance driven by a combination of spin Hall torque and field generated by the current \cite{Solyom2018}. The small signal is due to the minimal drive and readout efficiency occurring when the magnetization is oriented mostly along the in-plane hard axis. (The omnipresent features in the gray regions are artifacts arising from antenna modes of the waveguides and wirebond, and should be ignored.) 

With increased DC bias, spin-orbit torques (SOTs) from the platinum layer effectively antidamp the Py layer's magnetic motion, permitting excitation of large-amplitude, narrow-linewidth SW oscillation, and parametrically driven oscillations near twice the fundamental frequency (8 GHz) \cite{Solyom2018}. In this regime, evidence of two spin-wave modes (SW1 and SW2) can be resolved. A second (subtle) change in lineshape near and above 4.47 mA \emph{might} suggest the onset of DC-driven auto-oscillations phase-locked to the drive (similar to the changes observed in Ref.~\cite{Sankey2006}).

\section{ODMR measurements}

\begin{figure*}
\centering
\includegraphics[width=17.5cm]{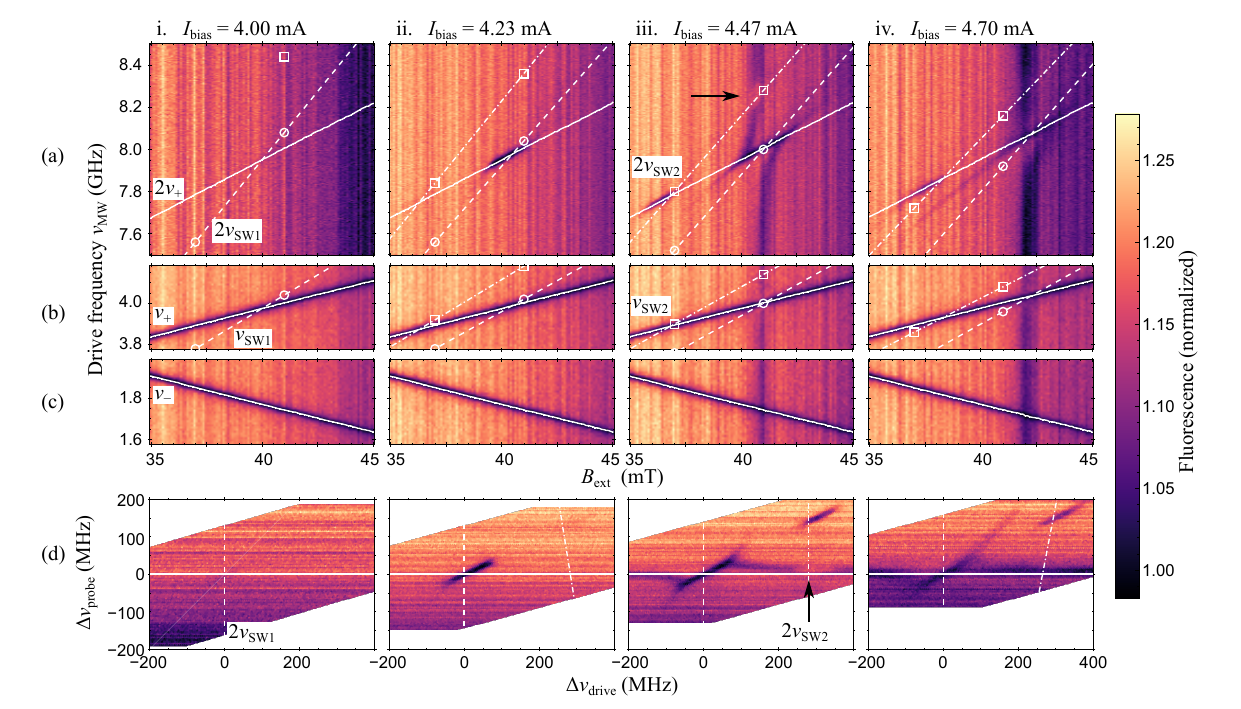}
\caption[ODMR measurements]{
\label{fig2:CW-ESR-data} 
\textbf{ODMR measurements.} 
(a-c) ODMR measurements of the probe N-$V$ center near the STO resonance. Circles (squares) represent the $\nu_\mathrm{SW1}$ ($\nu_\mathrm{SW2}$) mode frequency measured in transport (as shown in Fig. \ref{fig1:dev-transport}(c)), while the dash-dotted (dashed) lines are guides to the eye connecting the measurements. Solid white lines are the frequencies fit to the $\nu_{+}$ and $\nu_-$ N-$V$ transitions in rows (b) and (c), and represent $2 \nu_+$ in the row (a). Each vertical slice of data is normalized by the mixed-state fluorescence of the $\nu_{\pm}$ transitions. (d) Same data as in (a), but transformed into axes of probe vs drive detunings from the SW1 mode. The dotted diagonal line shows where $\Delta \nu_{\mathrm{probe}} = \Delta \nu_{\mathrm{drive}}$. Columns i-iv show the evolution of the measurement as the bias is swept from 4.0 to 4.7 mA. The black arrows in (a)iii and (d)iii show evidence of power sapping from the STO when parametrically driving SW2.}
\end{figure*}

In contrast to the electrical measurements above, the onset of auto-oscillation and phase locking presents a qualitatively distinctive feature in the ODMR spectrum of a single proximal N-$V$ spin. In this detection modality, the N-$V$ is continuously excited with green 532 nm laser light focused through a 0.95 NA air objective, while the resulting fluorescent emission (filtered by a 635 nm long pass) is collected by the same lens and monitored by a single-photon counter. The optical excitation polarizes the N-$V$ into the $m_s = 0$ spin state, which fluoresces more brightly than the $\pm1$ states. Then, any oscillating magnetic fields (coherent or noisy) at frequencies near the spin resonance can drive the N-$V$ into a spin mixture with reduced fluorescence.

Figure \ref{fig2:CW-ESR-data}(a)-(c) shows fluorescence data obtained by continuously applying a microwave current $I_{\mathrm{MW}} = 0.5$ mA through the wire while sweeping the applied field $B_\mathrm{ext}$ and microwave frequency $\nu_\mathrm{MW}$. Columns i-iv are the same data taken at varied $I_\mathrm{bias}$ to show the transition through the SW's critical current $\sim$4.4 mA. The frequency ranges in Fig. \ref{fig2:CW-ESR-data}(b)-(c) shows the ``standard'' ODMR response from driving the N-$V$ spin transition at frequencies of $\nu_\pm$ (the splitting of which increases with field due to the Zeeman effect) directly with the field from $I_\textrm{MW}$. Each vertical slice is simultaneously fit to two Lorentzian dips of the same width and depth to extract the frequencies $\nu_\pm$, as well as the magnitude of the (mixed-spin) fluorescence that occurs on resonance, which is subsequently used to normalize all vertical slices in Fig. \ref{fig2:CW-ESR-data} (this normalization compensates for slow drifts in laser power and alignment during data acquisition), including the data near the second harmonics in row (a). Note that the gradual reduction in fluorescence at higher $B_\mathrm{ext}$ arises from optically induced spin mixing associated with fields orthogonal to the N-$V$ near the excited-state level anti-crossing \cite{Tetienne2012}. Spin-wave frequencies $\nu_\mathrm{SW1}$ ($\nu_\mathrm{SW2}$) obtained by ST-FMR are superimposed as open circle (square) symbols and connected with guide lines in Fig. \ref{fig2:CW-ESR-data}(b), while the second harmonics of these same frequencies are plotted in Fig. \ref{fig2:CW-ESR-data}(a). The solid lines show $\nu_\pm$ and their harmonics for reference. In contrast to the magnetoresistive readout above, here we have reduced $I_\textrm{MW}$ to 0.5 mA so that it is insufficient to parametrically drive the SW modes with 4 mA of bias (Fig. \ref{fig2:CW-ESR-data}(a)i). 

Operating near the second harmonic offers an attractive method of probing for magnetic signatures, as the N-$V$ is insensitive to the direct drive from $I_\textrm{MW}$, while maintaining sensitivity to the SW stray fields near $\nu_\textrm{SW1,2}$. For example, as $I_\textrm{bias}$ approaches the STO threshold (ii), the expected feature appears in (a) at $B_\mathrm{ext} = 40$ mT where the parametrically driven SW1 and N-$V$ frequencies are in resonance \cite{Solyom2018}. As the bias is increased beyond the STO threshold (iii), several new features emerge. First, SW auto-oscillations directly drive the N-$V$ spin at 41 mT (independent of $I_\textrm{MW}$), when $\nu_\textrm{SW1}\approx \nu_+$, leading to a vertical stripe of reduced fluorescence. The stripe bends as the drive frequency approaches $2\nu_\mathrm{SW1}$ producing tails that extend well beyond the resonance condition. As discussed below, these features are reasonably captured by a simple macrospin model, but we can also anticipate the behavior qualitatively. At field strengths where $\nu_\textrm{SW1}=\nu_+$, i.e., when the drive frequency $\nu_\textrm{MW}$ approaches twice the STO resonance frequency, the magnetic oscillations can phase lock with the drive, pulling their frequency out of resonance with $\nu_+$ and returning the fluorescence to the higher $m_s=0$ value, e.g., near 8 GHz at 41 mT. When the phase-locked SW mode is driven to resonance with $\nu_\textrm{MW}=2\nu_+$ again, it drives N-$V$ transitions, and the fluorescence returns to the lower mixed-state value. At fields away from 41 mT, where $\nu_\textrm{SW1}\ne\nu_+$, the non-resonant drive \emph{modulates} the auto-oscillating spin wave, producing sidebands that can drive the NV transitions, generating the tails. This is most easily seen in Fig. \ref{fig2:CW-ESR-data}(d), which shows the same data as in (a), but transformed onto axes of N-$V$ ``probe'' detuning
\begin{align}
\Delta \nu_{\mathrm{probe}} & = \nu_+ - \nu_{\mathrm{SW}1}
\end{align}
and ``drive'' detuning
\begin{align}
\Delta \nu_{\mathrm{drive}} & = \nu_{\mathrm{MW}} - 2 \nu_{\mathrm{SW}1}
\end{align}
taken relative to $2 \nu_{\mathrm{SW}1}$, with an added dotted line along $\Delta\nu_\text{drive}=\Delta\nu_\text{probe}$ for reference. As expected for a second-harmonic drive, the parametrically driven and phase-locked dips in (ii) and (iii) occur at probe detunings of $\Delta \nu_{\mathrm{probe}} = \Delta \nu_{\mathrm{drive}}/2$. Above the STO threshold (iii and iv), we also see auto-oscillations as a horizontal line at $\Delta \nu_\mathrm{probe} = 0$, while the tails asymptote toward unity slope, consistent with $I_\text{MW}$ adding modulation sidebands to the SW oscillation. Between the phase-locked and asymptotic regime, the ODMR feature transitions continuously, indicating the drive may be pulling the oscillator frequency without fully phase-locking. We note that this spectral feature is suppressed in (iii) near the first harmonic (in (b)), as expected by symmetry for our geometry; the spin-orbit torques are nearly aligned with the average magnetization of the SW oscillations, and so directly driving its precession is inefficient.

In Fig.~\ref{fig2:CW-ESR-data}(a,iii), we also observe a second dip at $B_\mathrm{ext} = 36$ mT near the resonance condition $\nu_\text{SW2}=\nu_+$ for SW2. As such, we associate this feature with parametric driving of SW2. Interestingly, when the N-$V$ is resonant with the auto-oscillations of SW1 (41 mT), we observe a brightening of the fluorescence (arrow in (a,iii) and (d,iii)) while driving near $2\nu_\text{SW2}$, suggesting our parametric drive of SW2 has quenched the auto-oscillations of SW1.

Finally, as the bias is further increased (iv), we see the SW features darken and broaden, with the modulation tail visible over a greater range of fields. Additionally, the brightening associated with driving SW2 parametrically while probing SW1 is suppressed, suggesting that SW1 is no longer fully quenched.

\section{Macrospin modeling}

\begin{figure}[ht]
\includegraphics[width=8.5cm]{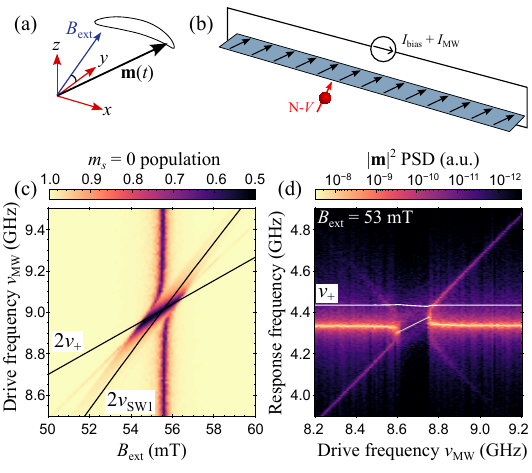}
\caption[Simulated ODMR]{
\label{fig3:Simulated-data} 
\textbf{Simulated ODMR.} 
(a) Axis definitions and typical large-amplitude trajectory of the magnetization $\mathbf{m}(t)$ in the macro-spin simulation. (b) The time-varying magnetic field at the N-$V$ probe location---which determines its population and fluorescence---arises from the magnetic layer's integrated dipolar field, the Oersted field from applied currents, and the external applied field. (c) Calculated $m_s = 0$ population when the nanowire is driven with $I_\mathrm{bias} = 4.47$ mA and $I_{\mathrm{MW}} = 0.5$ mA, which reproduces the STO ``vertical stripe'', phase locking region, and modulation sideband tails observed in Fig. \ref{fig2:CW-ESR-data}(a,iii). (d) Power spectral density (PSD) of the macro-spin model under the same conditions as (c), with a fixed magnetic field of $B_{\mathrm{ext}} = 53$ mT, clearly showing the phase-locked region from 8.6-8.75 GHz drive, frequency pulling near this region, and the modulation sidebands outside this region. The white line shows the N-$V$ frequency, which is slightly affected by the changing average magnetization.
}
\end{figure}

The signatures of parametric drive, auto-oscillations, and phase locking are reasonably captured by a simple model coupling macrospin (uniform) magnetization dynamics to an N-$V$ probe via the stray field. Specifically, the magnetization's unit vector $\mathbf{m}$ follows the Landau-Lifshitz equation
\begin{equation}\label{eq:LLSOT}
     \frac{d \mathbf{m}}{dt} = -\frac{\gamma}{\mu_0} \mathbf{m} \times \mathbf{B_{eff}} - \frac{\alpha}{\mu_0} \mathbf{m} \times (\mathbf{m} \times \mathbf{B_{eff}}) + \mathbf{q_{SOT}}
\end{equation}
with gyromagnetic ratio $\gamma = 28$ GHz/T, vacuum permitivity $\mu_0$, Gilbert damping coefficient $\alpha = 0.04$ (estimated from ST-FMR measurements \cite{Solyom2018}), plus an added spin-orbit torque (Fig. \ref{fig3:Simulated-data}(a))
\begin{align}
    \mathbf{q_{SOT}} & = \frac{\Theta_\text{SH}}{M_s L_y L_z^2} \frac{\mu_0 \gamma \hbar}{ 2 e} I_{\mathrm{Pt}} \hat{y}
\end{align}
with effective spin Hall angle $\Theta_\text{SH} = 0.1$ in the Pt layer (chosen so that the observed and modelled STO threshold currents match), Py saturation magnetization $M_s = 760$  \cite{Nibarger2003}, magnetic layer width $L_y=300$ nm and thickness $L_z=5$ nm, and Pt current
\begin{align}
    I_{\mathrm{Pt}} = \frac{\rho_{\mathrm{Py}}}{\rho_{\mathrm{Pt}} + \rho_{\mathrm{Py}}} I,
\end{align} 
where $\rho_{\mathrm{Pt}} = 21.9$ $\mathrm{\upmu\Omega \cdot cm}$ and $\rho_{\mathrm{Py}} = 65.2$ $\mathrm{\upmu\Omega \cdot cm}$  \cite{Duan2014Spin}. The first term in Eq.~\ref{eq:LLSOT} describes precession about an effective field
\begin{equation}
\mathbf{B_{eff}} = \mathbf{B_I} + \mathbf{B_{ext}} + \mathbf{B_{demag}} + \mathbf{B_{therm}},
\end{equation} 
where
\begin{equation}
    \mathbf{B_I} = \eta I \hat{y},
\end{equation}
is the average current-generated field experienced by $\mathbf{m}$ with efficiency $\eta=1$ T/A (roughly estimated from the geometry and resistivities of the metals, and consistent with the ST-FMR lineshape \cite{Solyom2018}),
\begin{equation}
\mathbf{B_{ext}} = B_\mathrm{ext} \left(\sqrt{\frac{2}{3}}\hat{y} + \frac{1}{\sqrt{3}}\hat{z} \right),
\end{equation}
is the externally applied field (magnitude $B_\text{ext}$),
\begin{equation}
\mathbf{B_{demag}} = -\mu_0 M_s \mathbf{N} \cdot \mathbf{m}
\end{equation}
is the demagnetizing field with (shape-defined) tensor $\mathbf{N}$, approximated here as that of an ellipsoid with only diagonal elements $N_{xx} = 0$, $N_{yy} = L_z/(L_y + L_z)$, and $N_{zz} = L_y/(L_y + L_z)$ \cite{Osborn1945, Solyom2018}, and $\mathbf{B_{therm}}$ is a stochastic Langevin field \cite{Brown1963Thermal} drawn from a Gaussian distribution of standard deviation $\sqrt{4\mu_0\alpha k_B T/\gamma M_s V \Delta t}$ at each time step (of duration $\Delta t$) and dimension separately, with Boltzmann constant $k_B$, temperature $T=300$ K, and magnetic volume $V=L_xL_yL_z= 4.5 \times 10^{-21}$ m$^3$ (estimated from the nominal Py dimensions). The second term in Eq.~\ref{eq:LLSOT} describes the magnetization's descent down the potential energy gradient due to damping $\alpha$. For each applied field and drive current, we allow the simulation to reach a steady state, then average the power spectral densities over 16 iterations of $T_{\mathrm{sim}} = 20$ $\upmu$s each.

Once the time-dependent magnetization is computed, the resulting N-$V$ dynamics are modeled from the local field at the position of the N-$V$, comprising the static applied field $\mathbf{B_\text{ext}}$, the Oersted field $\mathbf{B_\text{Oe}}$  from the current in both layers of the wire, and the dipolar field $\mathbf{B_\text{dip}}$ from the magnetic layer. The Oersted field is computed assuming translational symmetry along $\hat{x}$, while the dipolar field is obtained by assuming uniform magnetization of the Py along $\mathbf{m}$ and integrating over the layer's volume (see Fig. \ref{fig3:Simulated-data}(b)). We use $\mathbf{r_{NV}} = -(240, 300, 60)$ nm for the position of the N-$V$ probe relative to the center of the device, estimated from the optical measurements in Fig. \ref{fig1:dev-transport}(b) (determines $xy$) and SRIM calculations to determine the implantation depth \cite{Solyom2018}. The resulting time-dependent field at the N-$V$ is then used to calculate three relevant quantities: First, the N-$V$ spin transition frequency is set by the time-averaged field 
\begin{align}
    \langle B_j\rangle & = \frac{1}{T_{\mathrm{sim}}} \int_0^{T_{\mathrm{sim}}} \mathbf{B_\text{dip}} \cdot \mathbf{e}_j dt,
\end{align}
where $T_{\mathrm{sim}}$ is the simulation's duration and $\mathbf{e}_j$ are three orthogonal unit vectors
\begin{align}
    \mathbf{e}_X &= \hat{x}\\
    \mathbf{e}_Y &= \frac{1}{\sqrt{3}} (\hat{y} - \sqrt{2} \hat{z})\\
    \mathbf{e}_Z &= \frac{1}{\sqrt{3}} (\sqrt{2} \hat{y} + \hat{z})
\end{align}
reckoned relative to the N-$V$ symmetry axis $\mathbf{e}_Z$; second, the Rabi frequency
\begin{align}
    \Omega_R & = 4 \pi \gamma_{NV} \frac{\int_0^{T_{\mathrm{sim}}} \mathbf{B_\text{dip}} \cdot \mathbf{\sigma}(\nu_\mathrm{MW}/2) dt}{\int_0^{T_{\mathrm{sim}}} \mathbf{\sigma}(\nu_{\mathrm{MW}}/2) \cdot \mathbf{\sigma}(\nu_{\mathrm{MW}}/2) dt}
\end{align}
is calculated from coherent, synchronized magnetic oscillations at half the microwave drive frequency $\nu_\text{MW}$, where $\gamma_{NV} = 28$ GHz/T is the N-$V$ spin's gyromagnetic ratio and $\mathbf{\sigma}(\nu) = e^{i 2 \pi \nu t} \mathbf{e}_X + i e^{i 2 \pi \nu t} \mathbf{e}_Y$ is the transverse component of the spin's co-rotating frame; third, the induced relaxation rate
\begin{align}
    \Gamma_1 & = 4 \pi \gamma_{NV}^2 {T_{\mathrm{sim}}} \left|\frac{\int_0^{T_{\mathrm{sim}}} \mathbf{B_\text{dip}} \cdot \mathbf{\sigma}(\nu_{+}) dt}{\int_0^{T_{\mathrm{sim}}} \mathbf{\sigma}(\nu_{+}) \cdot \mathbf{\sigma}(\nu_{+}) dt} \right|^2
\end{align}
is found from the transverse field noise spectrum at the transition frequency $\nu_+$. With these parameters, we can estimate the steady state $m_s = 0$ population using a two-level optical Bloch equation \cite{Dreau2011} (see Appendix \ref{subsection: ODMR simulation}). 

By varying the applied field and drive frequency, we simulate data in Fig. \ref{fig3:Simulated-data}(c) over the same parameter range as the measurement in Fig. \ref{fig2:CW-ESR-data}(a,iii). For the chosen parameters, the simulation provides somewhat quantitative agreement with the key features in the measured ODMR spectrum, though there are some exceptions. First, the macrospin approximation precludes additional SW modes or associated interactions. Second, the simulated resonance between the N-$V$ and the magnet occurs at higher fields than experimentally observed, since the macrospin mode generally has lower frequency than a spatially varying mode that is ``stiffened'' by the exchange field. 

Despite this, our simple toy model validates the above interpretation of the ODMR signatures associated with phase-locking and freely-oscillating STO regimes, including frequency pulling near the phase locking regime, and a strong phase-locked response at $2\nu_+$. We also clearly see the modulation sidebands occurring at the expected frequencies. To make these features more explicit, Fig. \ref{fig3:Simulated-data}(d) shows the power spectral density (PSD) of the simulated macrospin (the summed contributions of $\mathbf{m}$'s $x$-, $y$-, and $z$-components), with the same axes as those of Fig.~\ref{fig2:CW-ESR-data}(d), but a color scale associated with $\mathbf{m}$ instead of N-$V$ fluorescence. At $\nu_\mathrm{MW}$ = 4.35 GHz, we observe the expected free-running STO, whose frequency is pulled down as the drive approaches 8.6 GHz, at which point the magnetization locks phase with the drive, following the frequency $\nu_\text{MW}/2$ until it unlocks at 8.75 GHz drive, where the STO is again free-running at a higher-than-unperturbed frequency. Modulation sidebands appear when the response (probe) and drive detunings are equal, as expected. As a caveat, note that this comparison is only rigorously valid outside the phase-locked regime, since, within it, the large Rabi frequency of the magnetization drive broadens the N-$V$'s ODMR linewidth rather than producting a delta-function frequency response expected from a coherent oscillation.

\section{Lifetime measurements}

\begin{figure}[ht]
\includegraphics[width=8.5cm]{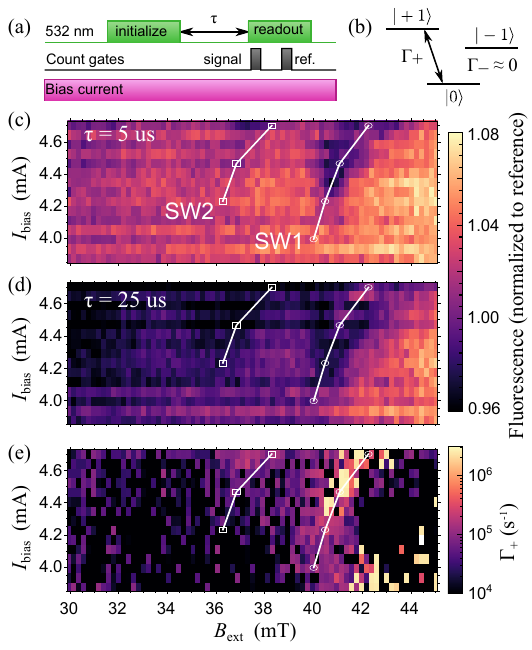}
\caption[Relaxation measurements]{
\label{fig4:Relaxometry data} 
\textbf{Relaxation measurements.} 
(a) Diagram of pulse measurement scheme used for relaxometry, showing the 532 nm laser pulses for initialization and readout, separated by delay $\tau$, with gated photon counting for the ``signal'' and ``reference'' times. Bias current is applied throughout. (b) Simplified population rates $\Gamma_\pm$ between the ground ($|0\rangle$) and two excited spin states ($|\pm 1\rangle$). (c), (d) Measured fluorescence during relaxometry for $\tau = 5$ $\upmu$s and $\tau = 25$ $\upmu$s, respectively. Fluorescence is normalized by the ``reference'' counts, and decreases when the the magnetic noise from a SW mode is resonant with the N-$V$. (e) Results to fitting a single-exponential decay model to the fluorescence data for $\tau \in  \{5, 10, 15, 20, 25\}$ $\upmu$s. Open circles (squares) denote the resonance conditions between the N-$V$ and SW1 (SW2) for (c), (d) and (e).
}
\end{figure}

Finally, we can study STO dynamics in the absence of microwave driving by performing all-optical relaxometry with the N-$V$ probe. Specifically, we perform lifetime measurements using the pulse sequence shown in Fig. \ref{fig4:Relaxometry data}(a), first initializing the N-$V$ into $m_s = 0$ with a 532 nm laser pulse, then allowing variable ``dark'' evolution time $\tau$ between 5 and 25 $\upmu$s, and finally reading out the new spin state with another laser pulse. The signal fluorescence is normalized to the counts measured during a later ``reference'' time to minimize slow drifts in the optical paths. A simplified single-decay model (Fig. \ref{fig4:Relaxometry data}(b)) is used to extract the relaxation rate $\Gamma_+$ assuming that the STO is the dominant source of relaxation during the evolution time and that it only drives relaxation between spin states 0 and +1.

Figures \ref{fig4:Relaxometry data}(c) and (d) show the fluorescence from the relaxation measurements with $\tau = 5$ and $25$ $\upmu$s as the external field and bias are swept. Open circles (squares) shown the bias and magnetic field conditions where the probe and spin-wave frequencies are in resonance $\nu_+ = \nu_{\mathrm{SW}1}$ ($\nu_+ = \nu_{\mathrm{SW}2}$). Immediately apparent are the dips in fluorescence that occur at biases near and above the STO thresholds. (Note that the apparent increase in fluorescence above $I_\mathrm{bias} = 4.5$ mA in Fig. \ref{fig4:Relaxometry data}(d) is an artifact of the normalization procedure: when $\Gamma_+$ exceeds the polarization rate during the laser pulse, the initialization fidelity is diminished, which reduces the reference counts, increasing the normalized signal.)

Next, we fit time-series data spanning $5~\upmu$s $< \tau <$ 25 $\upmu$s at each applied field and bias to a single-exponential decay in order to extract the relaxation rate $\Gamma_+$ shown in Fig. \ref{fig4:Relaxometry data}(e). We note that, while the all-optical relaxometry approach is experimentally simple to implement, parameter extraction from the fluorescence measurements is susceptible to crosstalk from other noise sources, such as spin mixing between $0$ and $-1$ states near the excited-state level anti-crossing at $B_{\mathrm{ext}} = 50.4$ mT during optical pumping. Furthermore, the restriction of $\tau \in [5, 25]$ $\upmu$s limits our ability to discern relaxation rates far outside the range 0.04 to 0.2 $\mathrm{\upmu s}^{- 1}$. Nevertheless, the simplistic approach to decay fitting is in agreement with our interpretation of the N-$V$ probing two spin-wave modes and confirms the STO threshold at $I_\mathrm{bias} = 4.3$ mA above which the measured magnetic-noise-induced relaxation rate increases by orders of magnitude. The measured value $\Gamma_+ > 1$ $\mathrm{\upmu s}^{- 1}$ when SW1 undergoes auto-oscillation are also consistent with the rates needed to form the features seen in ODMR, since the relaxation rates required to see any features in the latter must be comparable to the spin polarization rate of $\sim5$ MHz from optical pumping.

\section{Conclusion}

We report N-$V$ ODMR and spin relaxation measurements of a free-running spin torque oscillator (STO), and the ODMR spectral signatures of parametric phase-locking, near resonant frequency STO frequency pulling and modulation, and the quenching of one STO mode by another spin wave (SW). The interpretation of all but the last (multi-SW) feature are validated by a simple macrospin simulation. Beyond these demonstrations, pulsed ODMR might enable distinguishing between coherent phase locking and noisy auto-oscillations, and relaxometry protocols can be improved and expedited with MW spin preparation and adaptive pulse pattern generation \cite{Caouette2022}.

To maximize the utility of these testbed-system results, these techniques should be implemented in scanning probe systems, enabling one to resolve the spatial distribution of SW modes below the diffraction limit of laser light. In particular, this would complement and improve upon existing Brillouin light scattering techniques used to determine the localization of mutual synchronization in STO arrays \cite{Zahedinejad2019}. Even using standard ODMR, we demonstrate a measurement capable of observing STO dynamics far beyond what is possible with device-scale resistive readout.

\section*{Acknowledgements}

This work is partially supported by the National Science and Engineering Research Council (NSERC RGPIN 435554-13, RGPIN-2020-04095), Canada Research Chairs (229003 and 231949), Fonds de Recherche – Nature et Technologies (FRQNT PR-181274), the Canada Foundation for Innovation (Innovation Fund 2015 project 33488 and LOF/CRC 229003), and l’Institut Transdisciplinaire d’Information Quantique (INTRIQ). L. Childress is a CIFAR fellow in the Quantum Information Science program. A. Solyom acknowledges support from the NSERC CREATE program QSciTech. M. Caouette-Mansour acknowledges financial support from the First Nation Council of Innu Essipit. 

A.S. designed the experiment, acquired and analyzed the data, simulated the measurement, led the fabrication of the sample, and co-wrote the manuscript. M.C.-M. performed initial characterization of N-$V$s in proximity of candidate devices to select for microfabricating connections. B.R. developed the fabrication process for device connections. P.B. developed the subtractive fabrication process for patterning the nanowires. L.C. advised on experiments and simulations, and co-write the paper. J.S. advised on experiments and simulations, created the magnetic domain solver used for simulations, and co-wrote the paper.

\section*{Appendix}

\subsection{Device fabrication}\label{subsection:Device fab}

We use an electronic grade diamond (element6) with a (001) surface as the substrate for the nanofabricated device. The diamond has been implanted with a densely populated layer of N-$V$ centers created via $^{15}$N$^+$ ion implantation described in our previous experiment \cite{Solyom2018}. We fabricate the device in this study using a two-step process. To define the nanowires, 5 nm layers of permalloy and platinum are deposited by electron beam evaporation uniformly over the diamond surface, and a 10-nm-thick alumina mask in the shape of the nanowire is deposited by e-beam lithography and liftoff. The surrounding metal is then removed with an argon ion mill. We find this step greatly reduces the N-$V$ spin state fluorescence contrast, but that an oxygen plasma ``asher'' can recover the spin contrast without damaging the magnetic structures (see Appendix \ref{subsection: NV contrast recovery}). Finally, KOH is used to remove the alumina layer, allowing for top electrical contact. We then select devices with a well-coupled N-$V$s nearby, and optically re-image their positions relative to the nearby alignment marks (compensating for the systematic drift during their initial patterning, e.g., due to charging effects). Without this realignment step, deviations by up to 1 $\upmu$m from design specification precludes reliably creating lead overlaps of 0.3 $\upmu$m.


\subsection{N-$V$ ODMR contrast recovery}\label{subsection: NV contrast recovery}

\begin{figure}[ht]
\includegraphics[width=8.5cm]{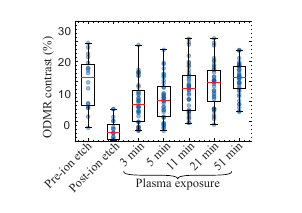}
\caption[ODMR contrast damage and recovery]{
\label{fig5:Appendix ODMR contrast recovery} 
\textbf{ODMR contrast damage and recovery.} 
Measured NV ODMR contrast for each process step. Each dot is a measurement of a fluorescent emitter's ODMR contrast, and the box plot shows the four quadrants in which the data lie. 
}
\end{figure}

Following the argon ion milling step above, we wirebond a stripline above the diamond approximately 100 $\upmu$m from the devices of interest and use this to generate microwave fields for testing nearby emitters. We noticed that ODMR contrasts from all N-$V$ centers were reduced by a factor of $\sim$8 (from $16\%$ to 2\%) as shown in Fig.~\ref{fig5:Appendix ODMR contrast recovery}, presumably from the ion mill step, consistent with an altered surface chemistry known to decrease the charge stability of the negatively charged N-$V$s and favor the magnetically-insensitive neutral state \cite{Petrakova2012, Fu2010}. To repair the damaged surface termination, we expose the patterned diamond surface to an oxygen plasma \cite{Favaro2015} from a plasma asher (Nanoplas DSB6000, 400 W RF power, 40 sccm O2 flow, 0.4 Torr chamber pressure at 45$^\circ$C). By wirebonding a stripline across the surface between each plasma exposure and measuring the ODMR contrast, we observe that the contrast measured over 40 emitters (not the same emitters as before the ion etch) recovers the nominal value after $\sim$50 minutes of exposure. Note the variation in observed contrast across the sampled N-$V$s may be partially due to variation in the local environments, as well as variation in the Rabi frequency, which was not controlled, as each emitter may have a different orientation relative to the microwave magnetic field.

While the plasma asher was shown to repair the measured spin contrast for the near-surface N-$V$s, we also verified that the processing was compatible with our fabrication flow. Specifically, we conducted AFM measurements of the alumina/Pt/Py trilayer patterned devices after each exposure, as well as after the alumina was removed by a KOH etch, in order to verify that the plasma did not etch the surface or patterned materials. Additionally, the ashing process was shown to not noticeably affect the magnetization of the Py (eg, by oxidization through the side walls) by measuring the splitting of N-$V$s spin resonances (partially set by the dipole field of the Py layer) at two locations near the nanowire after each step. At each step of the process, the so-inferred stray field from the Py was unchanged from the original value within our $\sim10\%$ measurement uncertainty.

\subsection{Simulation parameters of ODMR}\label{subsection: ODMR simulation}

Following the treatment by Dr\'{e}au et al \cite{Dreau2011}, we model the N-$V$ steady-state spin population under continuous wave excitation to simulate an analogue of the measured fluorescence. Specifically, we model the steady-state populations $p_0$ and $p_1$ of the $m_s$ = 0 and $m_s$ = 1 states respectively as

\begin{gather}
    p_1 = \frac{\Gamma_1 \left[ (2 \pi)^2(\nu_\mathrm{MW} - \nu_\mathrm{1})^2 + \Gamma_2^2 \right] + \Gamma_2 \Omega_R^2/2}{ (2 \Gamma_1 + \Gamma_p) \left[ (2 \pi)^2(\nu_\mathrm{MW} - \nu_\mathrm{1})^2 + \Gamma_2^2 \right] + \Gamma_2 \Omega_R^2}, \\
    p_0 = 1 - p_1.
\end{gather}
$\Omega_R$ and $\Gamma_1$ are the Rabi frequency and relaxation rate defined in the main text; $\Gamma_p = \Gamma_p^\infty \frac{s}{1+s}$ is the optically induced polarization rate and $\Gamma_2 = \Gamma_2^* + \Gamma_c^\infty \frac{s}{1+s} + \frac{1}{2} \Gamma_1$ is the spin dephasing rate. These in turn depend on the saturation parameter $s = 0.5$ (corresponding to the optical intensity used in these experiments), the optically induced polarization rate at saturation of $\Gamma_p^\infty = 5 \times 10^6$ s$^{-1}$, the optical cycling rate at saturation of $\Gamma_c^\infty = 8 \times 10^7$ s$^{-1}$, and the inhomogeneous dephasing rate $\Gamma_2^* = 2 \times 10^5$ s$^{-1}$.

As discussed in the main text, we decompose the time-dependent field from the simulated magnetization at the N-$V$'s position into left- and right-circulating drives, which we note can lead to double-counting of the spin flipping dynamics under some parameter ranges. Specifically, when the STO is phase-locked to the parametric drive, the magnet will drive coherent oscillations in the N-$V$ such that $\Omega_R > 0$, but if the drive frequency is exactly resonant with the N-$V$'s harmonic as $\nu_\mathrm{MW} = 2 \nu_1$ then the Fourier transform definition for $\Gamma_1 > 0$ also plays a role in mixing the spin. Because this double counting only occurs for values of $| \nu_{\mathrm{MW}} - 2\nu_1 | < \Delta \nu_\mathrm{FFT}$ (where $\Delta \nu_\mathrm{FFT} = 5$ MHz is the resolution of the Fourier transform), we neglect it, as it does not qualitatively change the results.

\subsection{Transport verification of the STO}

\begin{figure}
\vspace{0.5cm}
\includegraphics[width=8.5cm]{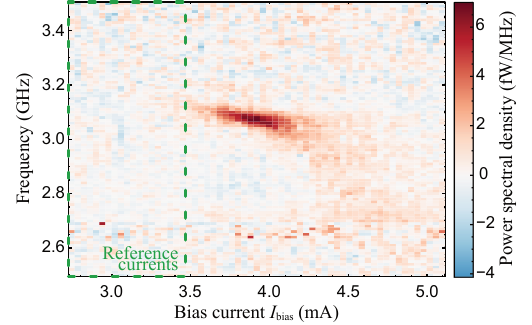}
\caption[Power spectral density of a spin torque oscillator]{
\label{fig6:Appendix PSD of STO} 
\textbf{Power spectral density of a second device.} 
Spectrum analyzer measurement of a STO-response from the same chip as the device studied in this paper. The power spectral density is shown relative to the measured values averaged over the reference currents shown in the green dashed box. 
}
\end{figure}

To confirm that the fabrication process leads to coherent, free-running STOs with signals detectable by conventional means, we use a spectrum analyzer having a resolution bandwidth of 22 MHz to measure the power spectral density of microwave electrical currents induced by the mixture of oscillating magnetoresistance with the bias currents \cite{Kiselev2003}. Although the device studied in this article was destroyed before we could perform this test, we measured the electrical spectrum in a device from the same chip and fabrication run with lateral dimensions of $6 \times 0.3$ $\upmu$m as shown in Fig. \ref{fig6:Appendix PSD of STO}. As the bias current is increased above the STO threshold of $\sim3.6$ mA, a spectral line with a with $<50$ MHz linewidth appears and decreases in frequency with bias, consistent with the macrospin model. This measurement further validates that the features in the N-$V$ fluorescence data can be attributed to the auto-oscillations of an STO. 

\bibliography{STO_synchronization}

\end{document}